\documentclass[10pt]{iopart}
\usepackage{graphicx}
\usepackage{bm}
\usepackage{amssymb}
\usepackage{mathrsfs}
\usepackage{color}

\def \Vec#1{\mbox{\boldmath $#1$}}
\def \p{\partial}
\def \f#1#2{\frac{#1}{#2}}
\def \mr#1{\mathrm{#1}}
\def \l{\left}
\def \ri{\right}

\def \dd#1#2{\frac{d#1}{d#2}} 
\def \pp#1#2{\frac{\p #1}{\p #2}} 
\def \Curl{\nabla\times} 

\def \nbl{\nabla}

\newtheorem{remark}{{\bf Remark}}

\newcommand{\qed}{\nobreak \ifvmode \relax \else
      \ifdim\lastskip<1.5em \hskip-\lastskip
      \hskip1.5em plus0em minus0.5em \fi \nobreak
      \vrule height0.75em width0.5em depth0.25em\fi}
      
\def \rmd{\mathrm{d}}

\begin{document}

\title{Relabeling symmetry in relativistic fluids and plasmas}

\author{Yohei Kawazura}
\address{Graduate School of Frontier Sciences, University of Tokyo, Kashiwanoha, Kashiwa, Chiba 277-8561, Japan}
\ead{kawazura@ppl.k.u-tokyo.ac.jp}

\author{Zensho Yoshida}
\address{Graduate School of Frontier Sciences, University of Tokyo, Kashiwanoha, Kashiwa, Chiba 277-8561, Japan}
\ead{yoshida@ppl.k.u-tokyo.ac.jp}

\author{Yasuhide Fukumoto}
\address{Institute of Mathematics for Industry Kyushu University, 744 Motooka, Nishi-ku, Fukuoka 819-0395, Japan}
\ead{yasuhide@imi.kyushu-u.ac.jp}
\begin{abstract}
The conservation of the recently formulated relativistic canonical helicity [Yoshida Z, Kawazura Y, and Yokoyama T 2014 \textit{J. Math. Phys.} \textbf{55} 043101] is derived from Noether's theorem by constructing an action principle on the relativistic Lagrangian coordinates 
(we obtain general cross helicities that include the helicity of the canonical vorticity).
The conservation law is, then, explained by the relabeling symmetry pertinent to the Lagrangian label of fluid elements.
Upon Eulerianizing the Noether current, the purely spatial volume integral on the Lagrangian coordinates is mapped to a space-time mixed three-dimensional integral on the four-dimensional Eulerian coordinates.
The relativistic conservation law in the Eulerian coordinates is no longer represented by any divergence-free current; 
hence, it is not adequate to regard the relativistic helicity (represented by the Eulerian variables) as a Noether charge, and this stands the reason why the ``conventional helicity'' is no longer a constant of motion.
We have also formulated a relativistic action principle of magnetohydrodynamics (MHD) on the Lagrangian coordinates, and have derived the relativistic MHD cross helicity.
\end{abstract}

\maketitle

%
%
\section{Introduction}\label{s:introduction}
The dynamics of an ideal fluid/plasma is constrained by infinitely many constants of motion.
Among them, the helicity dictates the invariance of the topology of vortex lines 
(in the present work, we assume a barotropic relation between the entropy and the temperature; then, the helicity is conserved in every co-moving fluid element confining the vortex lines).
According to Noether's earlier work~\cite{Noether}, a conservation law comes from a symmetry property.
Here, a symmetry denotes an invariance of action, or Lagrangian, for infinitesimal transformations (reparametrizations) of independent or/and dependent variables.
The primary example is the time reparametrization invariance eliciting the conservation of energy in classical mechanics.
In fluid/plasma theory, the symmetry leading the conservation of helicity is known as the ``relabeling symmetry'' pertinent to the Lagrangian labels of fluid elements (independent variables on Lagrangian coordinates)~\cite{Calkin,Salmon2,Yahalom,Padhye-Morrison1,Padhye-Morrison2,Fukumoto}.
Unlike the example of the time reparametrization, relabeling transformation is infinite dimensional, which resides in Noether's second theorem; 
see \cite{Olver} for extensive mathematical discussions around Noether's theorem and advanced topics, and~\cite{Newcomb,Bretherton,Ripa,Salmon1} for related applications of the relabeling symmetry.

The aim of this work is to examine this classical relation in the relativistic framework.
The reason why this practice interests us is because the conventional helicity is no longer a constant of motion in a relativistic fluid.
The space-time distortion (inhomogeneous Lorentz contraction due to non-constant velocity of the fluid) yields a ``relativistic baroclinic effect'' on a thermodynamically barotropic fluid, allowing a change in the circulation (or, equivalently, the vorticity)~\cite{MahajanYoshida2010}.
Yet, we can formulate a ``relativistic helicity'', by which we can delineate the topological constraint on the vortex lines in the relativistic space-time~\cite{YKY-JMP2014}.
As naturally expected, and as to be shown in Sec. \ref{s:R-Helicity conservation via Noether}, the conservation of the relativistic helicity is derived by the relabeling symmetry.
The key is to establish the `proper' correspondence between the Lagrangian frame and the Eulerian frame; the Noether current (being formulated by a Lagrangian formalism of the action) resides on the former, while the helicity is evaluated on the latter.  
Interestingly, the relativity reveals this fundamental relation with raising caution about the treatment of the proper-time in the Eulerian frame; 
the covector Eulerianized from the Noether current is not divergence-free, whereas it is divergence-free in the non-relativistic regime (so, it is often called a ``Noether current'' in the Eulerian frame).  

To deal with a charged fluid (plasma), we consider the canonical momentum that is the combination of the fluid momentum and the electromagnetic (EM) potential.
The (relativistic) helicity is defined for the canonical vorticity that is the curl (exterior derivative) of the canonical momentum.
Formulating the action principle on the Lagrangian frame, we can similarly relate the helicity to the Noether current (sec. \ref{ss:Noether PL}).
However, the complete action principle, encompassing the kinetic part producing the equation of motion and EM part producing Maxwell's equation, becomes a mixture of the Lagrangian kinetic term and the Eulerian EM term (we, then, need the protocol in evaluating the variation of the EM potential in the Lagrangian sense for deriving the equation of motion and in the Eulerian sense for deriving Maxwell's equation).
A solution to remove this inconvenience is to formulate the kinetic action in the Eulerian frame, but then, we miss the notion of the Lagrangian labels of fluid elements~\cite{Yoshida_Mahajan_PPCF2012}.
The action of the magnetohydrodynamics (MHD), however, can be formulated fully on the Lagrangian frame (Sec. \ref{ss:MHD Lagrange description}).
This is because, in the MHD model, the EM field (in fact, only the magnetic field is included as a state variable) is assumed to be co-moving with the fluid, and thus, Maxwell's equation is not needed.
The formulation of relativistic MHD action in Lagrangian coordinates is another new product of this work
(see~\cite{Dixon,Anile} for formulation of basic equations, \cite{Chiueh} for an Eulerian action principle, and \cite{Koide,Komissarov,Harikae} for applications in astrophysical computations).

This paper is organized as follows. 
In Sec. \ref{s:fundamentals}, we start with preliminaries of relativistic fluid, plasma and MHD. 
In Sec. \ref{s:R-Helicity}, we review the relativistic canonical helicity formulated in \cite{YKY-JMP2014}, 
and in Sec. \ref{s:Noether}, Noether's theorem applied for classical fields.
In Sec. \ref{s:Lagrange description}, the Lagrangian description of a relativistic fluid is given by following Salmon's formulation~\cite{Salmon3}, and formulate the actions of fluid, plasma, and MHD.
In Sec. \ref{s:R-Helicity conservation via Noether}, we derive the generalized relativistic helicity by Noether's theorem.
We also show the conservation of the relativistic cross helicity in MHD.

\begin{remark}[helicity and circulation]
\label{remark:circulation}
The invariance of the helicity implies various topological constraints on the field.
Let $\bm{a}$ be a three-dimensional field and $\bm{b}=\nabla\times\bm{a}$ (which is called he vorticity of $\bm{a}$).
For a fixed three-dimensional domain $\Omega$, which confines $\bm{b}$ (i.e., $\bm{n}\cdot\bm{b}=0$ on the
boundary $\partial\Omega$; $\bm{n}$ is the unit normal vector onto $\partial\Omega$),
the \emph{total helicity} is the volume integral $C=\int_\Omega \bm{a}\cdot\bm{b}\,\rmd^3x$,
which sums up the links, twists, and writes of all vortex lines confined in $\Omega$~\cite{Moffatt-Ricca-1992}.
When  $\bm{b}$ is ``filamentary'', $C$ evaluates the topological index of the filaments;
as the simplest setting, we may consider a pair of loops bounding disks, and define unit-vorticity filaments that are formally the delta-measures on the loops 
(see \cite{YKY-JMP2014} for a mathematical formulation in the context of Banach algebra), 
and then, $C$ evaluates the Gauss linking number of the loops.
A \emph{local helicity} can be defined by introducing a ``co-moving'' volume $W(t)$
that moves with a velocity $\bm{V}$, and integrating
$C_W=\int_{W(t)}  \bm{a}\cdot\bm{b}\,\rmd^3x$.
Here $W(t)$ is co-moving with $\bm{b}$ in the sense that $\bm{b}$ is transported by the
same velocity $\bm{V}$, i.e., $\bm{b}$ (2-form) satisfies $\partial_t\bm{b} -\nabla\times(\bm{V}\times\bm{B}) =0$.
If  $\bm{n}\cdot\bm{b}=0$ on $\partial W(t)$, $C_W$ is a constant of motion,
The relativistic helicity\,\cite{YKY-JMP2014} is defined by generalizing these relations for the
four-dimensional Minkowski space-time; the vorticity is, then, the exterior derivative $\rmd a$ of
a 1-form $a$, and the integrand $\bm{a}\cdot\bm{b}$ is identified to be the 3-form $a\wedge \rmd a$,
which is integrated over a co-moving 3-chain $W(t)$ embedded in the four-dimensional space-time
(which is no loner purely spatial); see Sec.\,\ref{s:R-Helicity} for a short review.

Introducing an arbitrary co-moving exact 2-form $\bm{c}$ (which may not be a physical quantity, and then we call it a \emph{mock field}, cf.\,\cite{YM_FDR2014}),
we can define a constant of motion $C_c=\int_{W(t)} \bm{a}\cdot\bm{c}\,\rmd^3x$,
which is called a \emph{cross helicity}\,\cite{Fukumoto}.
When $\bm{c}$ is a filament on a co-moving loop $L(t)$, $C_c$ evaluates the \emph{circulation} 
$\oint_{L(t)}\bm{a}\cdot\bm{\tau}\,\rmd x$, where $\bm{\tau}$ is the unit tangential vector on $\Gamma(t)$.

In the context of Hamiltonian mechanics (Eulerian representation), 
a helicity is a Casimir invariant pertinent to the degeneracy (or noncanonicality) of the Poisson bracket\,\cite{Morrison_RMP1998}.
However, a general topological invariant is not necessarily a Casimir invariant;
the circulation is such an example.
We may yet define a \emph{cross helicity} by associating the invariant with a co-moving \emph{mock field},
which is a Casimir invariant in the extended phase space, 
i.e., the product space of the original Poisson manifold and the mock field\,\cite{YM_FDR2014}.
The co-moving mock field corresponds to the vector generating the relabeling group\,\cite{Moreau}.
\end{remark}

%
%
\section{Preliminaries of relativistic fluid, plasma and MHD}\label{s:fundamentals}
We use the notation of Minkowski space time.
The reference-frame coordinates are denoted by
\begin{eqnarray*}
	x^\mu := (ct,x,y,z), \quad x_\mu := (ct,-x,-y,-z),
\end{eqnarray*}
where $c$ is the speed of light.
The Minkowski metric tensor is $g_{\mu\nu} = \mr{diag}(+, -, -, -)$.
The gradients are denoted as $\p_\mu = \p/\p x^\mu$ and $\p^\mu = \p/\p x_\mu$.
The relativistic 4-velocity is defined as
\begin{eqnarray*}
	{U}^\mu := \f{dx^\mu}{ds} = \l( \gamma ,\, \gamma\f{\bm{v}}{c} \ri), \quad
	{U}_\mu := \f{dx_\mu}{ds} = \l( \gamma ,\, -\gamma\f{\bm{v}}{c} \ri),
\end{eqnarray*}
where $s$ is the proper time, $\bm{v}$ is the reference-frame 3-velocity, and $\gamma = 1/\sqrt{1 - (v/c)^2}$ is the Lorentz factor.

\subsection{Fluid}
We start by introducing the thermodynamic enthalpy $h = mc^2 + {\mathcal E}( n,\, \sigma) + p/n$, 
where $m$ is the rest mass of a particle, ${\mathcal E}$ is the internal energy, $n$ is the rest frame particle density, $\sigma$ is the specific entropy and $p$ is pressure.
Then the 4-momentum is defined as $P_\mu := hU_\mu$.
The exterior derivative of the 4-momentum, $M_{\mu\nu} := \p_\mu P_\nu - \p_\nu P_\mu$, is a vorticity field tensor.
The equation of motion is given as 
\begin{equation}
	U^\mu M_{\mu\nu} = -\p_\nu \theta.
\label{e:Eulerian e.o.m. FL}
\end{equation}
Here we have assumed a barotropic relation $Td\sigma = d\theta$ ($T$ : temperature, $\theta$ : some function of $ n$).
The thermodynamic first law is, then, 
\begin{equation}
	dh = d\theta + \f{dp}{ n}.
\label{e:thermodynamic relation}
\end{equation}
Due to the barotropicity the internal energy is written as ${\mathcal E} = {\mathcal E}(n)$.
The dual of $M_{\mu\nu}$ is defined as
\begin{equation}
	M^{*\mu\nu} := \f{1}{2}\epsilon^{\mu\nu\alpha\beta}M_{\alpha\beta},
\label{e:def M* FL}
\end{equation}
where $\epsilon^{\mu\nu\alpha\beta}$ is the four dimensional Levi-Civita symbol.
We have the identity
\begin{equation}
	\p_\mu M^{*\mu\nu} = 0.	
\label{e:div M FL}
\end{equation}
Differentiating (\ref{e:Eulerian e.o.m. FL}), we obtain the vorticity equation
\begin{equation*}
	\p_\lambda(U^\mu M_{\mu\nu}) - \p_\nu(U^\mu M_{\mu\lambda}) = 0.
\end{equation*}
Substituting (\ref{e:def M* FL}), we obtain
\begin{equation}
	\p_\lambda(U^\lambda M^{*\mu\nu}) = M^{*\lambda\nu}\p_\lambda U^\mu + M^{*\mu\lambda}\p_\lambda U^\nu.
\label{e:Eulerian vorticity eq.2 FL}
\end{equation}

\subsection{Charged fluid (plasma)}
When the fluid particles are charged, we have to dress the momentum by the electromagnetic potential, and consider the ``canonical momentum'',
\begin{eqnarray*}
	{\mathcal P}_\mu := P_\mu + eA_\mu,
\end{eqnarray*}
where $e$ is the charge and $A_\mu$ is the 4-potential.
We write the Faraday tensor as $F_{\mu\nu} = \p_\mu A_\nu - \p_\nu A_\mu$.
The dual of $F_{\mu\nu}$ is defined in the same way as (\ref{e:def M* FL}).
Electric field and magnetic field are given as $\bm{E} = -\nbl A^0 - c^{-1}\bm{A}$ and $\bm{B} = \Curl \bm{A}$.
We have the identity
\begin{equation}
	\p_\mu F^{*\mu\nu} = \p_\mu F_{\nu\lambda} + \p_\nu F_{\lambda\mu} + \p_\lambda F_{\mu\nu} = 0.
\label{e:div F*EM}
\end{equation}
Here we consider a single species plasma, but if the plasma consists of multiple fluids, we have to sum the 4-currents of all species.
We have
\begin{eqnarray}
	U^\mu{\mathcal P}_\mu = h + e\varrho \quad (\varrho := U^\mu A_\mu).
\label{e:UP}
\end{eqnarray}
The vorticity field tensor is extended as a \emph{canonical} vorticity field tensor ${\mathcal M}_{\mu\nu} := \p_\mu {\mathcal P}_\nu - \p_\nu {\mathcal P}_\mu$.
The equation of motion is given as
\begin{equation}
	U^\mu {\mathcal M}_{\mu\nu} = -\p_\nu \theta.
\label{e:Eulerian e.o.m. PL}
\end{equation}
Equations (\ref{e:div M FL}) and (\ref{e:Eulerian vorticity eq.2 FL}) are modified by replacing $M$ by ${\mathcal M}$:
\begin{equation}
	\p_\mu {\mathcal M}^{*\mu\nu} = 0,
\label{e:div M PL}
\end{equation}
and
\begin{equation}
	\p_\lambda(U^\lambda {\mathcal M}^{*\mu\nu}) = {\mathcal M}^{*\lambda\nu}\p_\lambda U^\mu + {\mathcal M}^{*\mu\lambda}\p_\lambda U^\nu.
\label{e:Eulerian vorticity eq.2 PL}
\end{equation}

\subsection{MHD}
Next we review the MHD equation~\cite{Dixon,Anile}.
We define
\begin{equation*}
	\tilde{E}_\nu := U^\mu F_{\mu\nu} = \l(\gamma \Vec{E}\cdot\l( \f{\Vec{v}}{c} \ri),\; -\gamma\l( \Vec{E} + \f{\Vec{v}}{c}\times\Vec{B} \ri)\ri),
\end{equation*}
\begin{equation*}
	\tilde{B}^\nu := U_\mu F^{*\mu\nu} = \l(-\gamma \Vec{B}\cdot\l( \f{\Vec{v}}{c} \ri),\; -\gamma\l( \Vec{B} - \f{\Vec{v}}{c}\times\Vec{E} \ri)\ri).
\end{equation*}
The 3-vector parts of $\tilde{E}_\nu$ and $\tilde{B}^\nu$ correspond to the Lorentz transformations of $\bm{E}$ and $\bm{B}$, respectively.
In MHD, we assume
\begin{equation}
	\Vec{E} + (\Vec{v}/c)\times\Vec{B} = 0,
\label{e:Ohm's law}
\end{equation}
which is equivalent to $\tilde{E}_\nu = 0$.
We eliminate $\bm{E}$ from $F^{*\mu\nu}$ by using (\ref{e:Ohm's law}) and define   
\begin{equation*}
	{\mathcal F}^{*\mu\nu} := F^{*\mu\nu}|_{\bm{E} = \bm{B}\times\bm{v}/c},
\end{equation*}
which we call the ``MHD tensor''.
As (\ref{e:div F*EM}), we have
\begin{equation}
	\p_\mu {\mathcal F}^{*\mu\nu} = 0.
\label{e:div F}
\end{equation}
We denote
\begin{equation*}
	b^\nu := \tilde{B}^\nu|_{\bm{E} = \bm{B}\times\bm{v}/c} = U_\mu {\mathcal F}^{*\mu\nu} = \l(-\f{\gamma}{c}\Vec{v}\cdot\Vec{B},\; -\f{1}{\gamma}\Vec{B} - \f{\gamma}{c^2}(\Vec{v}\cdot\Vec{B})\Vec{v}\ri).
\end{equation*}
The norm of $b^\nu$ is 
\begin{equation*}
	b^2 = b^\nu b_\nu = -\f{|\bm{B}|^2}{\gamma^2} - \f{(\bm{v}\cdot\bm{B})^2}{c^2}.
\end{equation*}
At the non-relativistic limit ($|\bm{v}/c| \ll 1$), $b^\nu$ becomes $(0,\, -\bm{B})$. 
We remark that $b^\nu$ is \emph{orthogonal} to $U_\nu$:
\begin{eqnarray*}
	U_\nu b^\nu = 0,
\end{eqnarray*}
and the 4-dimensional divergence of $b^\nu$ is not zero:
\begin{equation}
	\p_\nu b^\nu = {\mathcal F}^{*\mu\nu}\p_\nu U_\mu = -b^\mu U^\nu \p_\nu U_\mu.
\label{e:div b}
\end{equation}
We may retrieve ${\mathcal F}^{*\mu\nu}$ as
\begin{eqnarray*}
	{\mathcal F}^{*\mu\nu} = U^\mu b^\nu - U^\nu b^\mu.
\end{eqnarray*}
In the place of the vorticity equation (\ref{e:Eulerian vorticity eq.2 FL}), we have
\begin{equation*}
	\p_\lambda(U^\lambda {\mathcal F}^{*\mu\nu}) = {\mathcal F}^{*\lambda\nu}\p_\lambda U^\mu + {\mathcal F}^{*\mu\lambda}\p_\lambda U^\nu.
\end{equation*}

The energy-momentum tensor of MHD is given as
\begin{equation*}
	T^{\mathrm{MHD}}_{\mu\nu} =  n h U_\mu U_\nu - pg_{\mu\nu} - b^2U_\mu U_\nu + \f{b^2}{2}g_{\mu\nu} - b_\mu b_\nu.
\end{equation*}
The equation of motion, $\p^\mu T^{\mathrm{MHD}}_{\mu\nu} = 0$, reads
\begin{equation}
	U^\mu M_{\mu\nu} = -\p_\nu \theta + \f{1}{ n}\l[ \p^\mu(b^2U_\mu U_\nu + b_\mu b_\nu) - \p_\nu\l( \f{b^2}{2} \ri) \ri].
\label{e:Eulerian MHD e.o.m.}
\end{equation}
The last term on the right-hand side is the Lorentz force.

\subsection{Special Relativistic Helicity}\label{s:R-Helicity}
The 4 velocity $U$ generates a diffeomorphism group $T(s)$, where $s$ is the proper time.
Let $V_0$ be a spatial volume. We denote $V(s) = T(s)V_0$.
The relativistic helicity~\cite{YKY-JMP2014} is
\begin{eqnarray*}
	\int_{V(s)} {\mathcal P} \wedge {\mathcal M} &=& \int_{V(s)} \f{1}{2}{\mathcal P}_\nu {\mathcal M}_{\alpha\beta}dx^\nu \wedge dx^\alpha \wedge dx^\beta \nonumber \\
													&=& \int_{V(s)} {\mathcal P}_\nu {\mathcal M}^{*\mu\nu}dV_\mu,
\end{eqnarray*}
where $dV_\mu$ is the Hodge dual of $-dx_\mu$, which is shown to be invariant if the barotropic equation of motion~(\ref{e:Eulerian e.o.m. PL}) holds, i.e.
\begin{equation}
	\dd{}{s}\int_{V(s)}{\mathcal P} \wedge {\mathcal M} = \int_{V(s)}L_U({\mathcal P}\wedge {\mathcal M}) = 0,
	\label{e:canonical helicity conservation}
\end{equation}
where $L_U$ denotes a Lie derivative along $U$.
Putting $e = 0$ in ${\mathcal P}$ yields the conservation of the helicity of a neutral fluid ($\int_{V(s)}P \wedge M$), and $m = 0$, the magnetic helicity ($\int_{V(s)}A \wedge d{\mathcal F}$) conservation in MHD;
(\ref{e:Ohm's law}) is the massless limit of~(\ref{e:Eulerian e.o.m. PL}) and $d\theta = 0$.
We note that ${\mathcal M}$ in (\ref{e:canonical helicity conservation}) may be replaced by an arbitrary exact 2-form ${\mathcal W}$ that obeys
\begin{eqnarray*}
	L_U{\mathcal W} = di_U{\mathcal W} = 0,
\end{eqnarray*}
where $i_U$ is a interior product with $U$, and then, $\int_{V(s)}{\mathcal P} \wedge {\mathcal M}$ is a cross helicity.
The invariance of this cross helicity is a straight forward generalization of Theorem 1 of~\cite{YKY-JMP2014}.
In the case of MHD, however, the cross helicity is special $\int_{V(s)}P \wedge {\mathcal F}$, which is initially discovered in this work.

%
%
\section{Noether's theorem}\label{s:Noether}
In this section we briefly review Noether's theorem for Lagrangian mechanics of general field.
We consider a first-order action such that
\begin{equation*}
	S = \int_D {\mathcal L}(q^\nu(a),\: \tilde{\p}_\mu q^\nu(a),\: a^\mu)d^4 a,
\end{equation*}
where ${\mathcal L}$ is the Lagrangian density, $a^\mu$ is the Lagrangian coordinates, $q^\nu(a)$ is the field variable, $\tilde{\p}_\mu = \p/\p a^\mu$ (the tilde is used to distinguish from the derivative with respect to $x^\mu$ in the previous sections) and $D$ is the domain.

The equation of motion (Euler-Lagrange equation) is assumed to be satisfied:
\begin{equation}
	\pp{\mathcal{L}}{q^\nu} - \tilde{\p}_\mu\l( \pp{\mathcal{L}}{(\tilde{\p}_\mu q^\nu)} \ri) = 0.
\label{e:e.o.m}
\end{equation}
Then we consider a transformation 
\begin{eqnarray*}
	a^\mu  &\mapsto& a'^\mu = a^\mu + \delta a^\mu. 
\end{eqnarray*}
The field variable is altered as
\begin{eqnarray*}
	q^\nu(a) &\mapsto& q'^\nu(a') = q^\nu(a) + \delta q^\nu(a) + \delta a^\lambda\tilde{\p}_\lambda q^\nu(a) =: q^\nu(a) + \Delta q^\nu(a).
\end{eqnarray*}
The change $\Delta q^\nu(a)$ is sometimes called ``total variation''; the first term is the change of $q$ on fixed $a$, and the second term is caused by the variation of $a$.
The change of the action under these transformations is calculated as,
\begin{eqnarray}
	\delta S = \int_{D'} \mathcal{L}(q'^\nu(a'),\: \tilde{\p}'_\mu q'^\nu(a'),\: a'^\mu) d^4a' - \int_D \mathcal{L}(q^\nu(a),\: \tilde{\p}_\mu q^\nu(a),\: a^\mu) d^4a \nonumber\\
	\nonumber\\
					 = \int_D\l\{ \delta q^\nu\l[ \pp{\mathcal{L}}{q^\nu} - \tilde{\p}_\mu\l( \pp{\mathcal{L}}{(\p_\mu q^\nu)} \ri) \ri] + \tilde{\p}_\mu\l[ \mathcal{L}\delta a^\mu + \delta q^\nu\pp{\mathcal{L}}{(\tilde{\p}_\mu q^\nu)} \ri]  \ri\}d^4a. \nonumber\\
\label{e:delta S}
\end{eqnarray}
where we used $\Delta\tilde{\p}_\mu q^\nu  = \tilde{\p}_\mu\Delta q^\nu - (\tilde{\p}_\mu\delta a^\lambda)\tilde{\p}_\lambda q^\nu$. 
The first term in the integrand disappears because of the equation of motion (\ref{e:e.o.m}).
When the integrand of the right-hand side is written as $\tilde{\p}_\mu \delta \Lambda^\mu$ where $\delta \Lambda^\mu$ is some function, the transformation $\delta a$ is called invariant transformation.
For such transformations, action invariance $\delta S = 0$ induces four dimensional divergence free current ($\tilde{\p}_\mu I^\mu = 0$), so-called Noether current:
\begin{equation*}
	I^\mu := \mathcal{L}\delta a^\mu + \delta q^\nu\pp{\mathcal{L}}{(\tilde{\p}_\mu q^\nu)} - \delta \Lambda^\mu.
\end{equation*}

%
%
\section{Lagrange description of fluid, plasma and MHD}\label{s:Lagrange description}

\subsection{Fluid}
We show the Lagrangian description of relativistic fluid which was first proposed by Salmon~\cite{Salmon3}, then extend it to plasma and MHD.
The Lagrangian description of the relativistic plasma and MHD is firstly proposed in this paper.
Let $q^\mu(a) = (t, \bm{x})$ be Eulerian coordinates and $a^\mu = (s, \bm{a})$ be Lagrangian coordinate, where $t$ and $s$ represent reference and proper time respectively.
The relativistic 4-velocity is written as
\begin{equation*}
	U^\mu(x) := \l.\dd{q^\mu}{s}\ri|_{a = q^{-1}(x)} = \dot{q}^\mu|_{a = q^{-1}(x)}.
\end{equation*}
We define the Jacobian between the Lagrangian and the Eulerian coordinates $J := \p(x)/\p(a)$ and $R := \sqrt{\dot{q}^\mu\dot{q}_\mu}$.
Then the proper number density is given as
\begin{equation*}
	 n(t,\,\bm{x}) = \f{R n_0(\bm{a})}{J},
\end{equation*}
where $ n_0$ is the initial rest number density.
We define $C_\mu^{\;\;\nu}$ as the cofactor of the matrix element $\p q^\mu/\p a^\nu$.
The identities of $J$ and $C_\mu^{\;\;\nu}$ which will be used repeatedly in the following calculations are provided in the appendix.

Now, the action is written as
\begin{eqnarray}
	S = \int \Bigl\{- n[mc^2 + {\mathcal E}( n)]\Bigr\} d^4x &=& \int \l\{-R n_0\l[mc^2 + {\mathcal E}\l(\f{R n_0}{J}\ri)\ri]\ri\} d^4a \nonumber\\ 
	&=:& \int {\mathcal L}_\mathrm{FL}d^4a,
\label{e:FL Lagrangian}
\end{eqnarray}
where ${\mathcal E}( n)$ is the barotropic internal energy.
Although it is obvious that $R = 1$, this evaluation is done after taking the variation of the action. 
Let us consider the variation $\delta q^\nu$.
We find
\begin{equation*}
	\delta S = \int \l[  - n_0h\dot{q}_\nu\delta\dot{q}^\nu + pC_\nu^{\;\;\mu}\pp{(\delta q^\nu)}{a^\mu}  \ri]d^4a.
\end{equation*}
where $p =  n^2\p {\mathcal E}/\p n$ is used.
The invariance of the action leads the equation of motion as
\begin{equation}
	 n_0\pp{}{s}(h\dot{q}_\nu) - C_\nu^{\;\;\mu}\pp{p}{a^\mu} = 0.
\label{e:Lagrangian e.o.m.}
\end{equation}
Converting to Eulerian coordinate, we obtain 
\begin{equation*}
	U^\mu\p_\mu P_\nu - \f{\p_\nu p}{ n} = 0,
\end{equation*}
where we used the formula (\ref{e:A d/da = J d/dx}). This is equivalent to (\ref{e:Eulerian e.o.m. FL}). 

\subsection{Plasma}
We modify the action (\ref{e:FL Lagrangian}) so as to include the interaction with EM field
\begin{eqnarray}
	S &= & \int \l\{-R n_0\l[mc^2 + {\mathcal E}\l(\f{R n_0}{J}\ri)\ri] - n_0e A_\nu(q)\dot{q}^\nu\ri\} d^4a \nonumber \\ 
	  &=:& \int {\mathcal L}_\mathrm{PL}d^4a.
\label{e:PL Lagrangian}
\end{eqnarray}
The variation $\delta q$ induces the change of the action as
\begin{eqnarray}
	\delta S = \nonumber\\
	\int \l[  - n_0(h\dot{q}_\nu + eA_\nu)\delta\dot{q}^\nu + pC_\nu^{\;\;\mu}\pp{(\delta q^\nu)}{a^\mu} - n_0 e \dot{q}^\mu\pp{A_\mu}{q^\nu}\delta q^\nu  \ri]d^4a.
\label{e:variation of PL action}
\end{eqnarray}
The equation of motion is obtained by $\delta S = 0$ and converting it to Eulerian coordinate, we get 
\begin{equation*}
	U^\mu\p_\mu {\mathcal P}_\nu - \f{\p_\nu p}{ n} - eU^\mu\p_\nu A_\mu = 0.
\end{equation*}
This is equivalent to (\ref{e:Eulerian e.o.m. PL}). 

To obtain Maxwell's equations, we need to add the EM Lagrangian, then the action is
\begin{equation*}
	S = \int \l[- n(mc^2 + {\mathcal E}) - neA_\mu U^\mu - \f{1}{4}F^{\mu\nu}F_{\mu\nu} \ri] d^4x.
\end{equation*}
We take variation with respect to $A_\nu$.
The careful treatment is required for this action.
As we showed above, we must not consider the EM Lagrangian for the derivation of the equation of motion.
On the other hand, for the derivation of Maxwell equation, the variation need to be carried out in Eulerian coordinate.
This ad hoc treatment is known for the particle mechanics~\cite{Landau}.

\subsection{MHD}\label{ss:MHD Lagrange description}
Next we derive Lagrangian description of MHD.
Before looking into the action, we investigate Lagrangian description of magnetic field like variable $b^\mu$.
Let us start by manipulating the MHD tensor as
\begin{eqnarray*}
	{\mathcal F}^{*\mu\nu}	&=& \l. \epsilon^{\mu\nu\alpha\beta}\pp{ A_\beta}{x^\alpha} \ri|_{\bm{E} = \bm{B}\times\bm{v}/c} = \l. \f{\epsilon^{\lambda\kappa\gamma\delta}}{J}\pp{q^\mu}{a^\lambda}\pp{q^\nu}{a^\kappa}\pp{ A_\beta}{a^\gamma}\pp{q^\beta}{a^\delta} \ri|_{\bm{E} = \bm{B}\times\bm{v}/c} \\
	\\
	\\
	            &=& \l[ \f{\epsilon^{0ijk}}{J}\l( \dot{q}^\mu\pp{q^\nu}{a^i} - \dot{q}^\nu\pp{q^\mu}{a^i} \ri)\pp{ A_\lambda}{a^j}\pp{q^\lambda}{a^k} \ri. \\
							& &\hspace{5em} \l.+ \f{\epsilon^{0ijk}}{J}\pp{q^\mu}{a^i}\pp{q^\nu}{a^j}\l( \dot{ A}_\lambda\pp{q^\lambda}{a^k} - \dot{q}^\lambda\pp{ A_\lambda}{a^i} \ri) \ri]_{\bm{E} = \bm{B}\times\bm{v}/c},
\end{eqnarray*}
where we used the formula (\ref{e:J epsilon}) in the second equality and the Latin indices represent $1,\,2,\,3$.
The second term in the right-hand side equals to $U^\mu F_{\mu\lambda}(\p q^\lambda/\p a^k)|_{\bm{E} = \bm{B}\times\bm{v}/c} = \tilde{E}_\lambda(\p q^\lambda/\p a^k)|_{\bm{E} = \bm{B}\times\bm{v}/c} = 0$.
Then we define the vector $b_0^i$ on Lagrangian coordinates as
\begin{equation*}
	b_0^i := \epsilon^{0ijk}\pp{ A_\lambda}{a^j}\pp{q^\lambda}{a^k}.
\end{equation*}
Defining ${\mathscr A}_i :=  A_\mu\p q^\mu/\p a^i$ and $\nbl_a$ as a gradient in $a$ space, $b_0^i$ is expressed as
\begin{eqnarray*}
	b_0^i = (\nbl_a \times \bm{{\mathscr A}})^i.
\end{eqnarray*}
We can show $b_0^i$ satisfies
\begin{equation}
	\pp{ b_0^i}{a^i} = 0 \;\;\;\;\mathrm{and}\;\;\;\; \dot{b_0^i} = 0.
\label{e:div b0 and dot b0}
\end{equation}
Now ${\mathcal F}^{*\mu\nu}$ is rewritten as
\begin{equation*}
	{\mathcal F}^{*\mu\nu} = \f{Rb_0^i}{J}\l( \dot{q}^\mu\pp{q^\nu}{a^i} - \dot{q}^\nu\pp{q^\mu}{a^i} \ri).
\end{equation*}
Therefore $b^\nu$ is rewritten using $ b_0^i$ as
\begin{equation*}
	b^\nu = U_\mu {\mathcal F}^{*\mu\nu} = \f{Rb_0^i}{J}\l( \pp{q^\nu}{a^i} - \dot{q}^\nu\dot{q}_\mu\pp{q^\mu}{a^i} \ri).
\end{equation*}
This tells us that $b^\nu$ is obtained by multiplying $(R/J)(\p q_\mu/\p a^i)$ (Eulerianization) and $g^{\mu\nu} - U^\mu U^\nu$ (projector orthogonal to $U^\mu$) to $b_0^i$.

Now we write the action of MHD by adding $b^2/2$ to (\ref{e:FL Lagrangian}) as 
\begin{eqnarray}
	S &=& \int \l\{- n[mc^2 + {\mathcal E}( n)] + \f{b^2}{2} \ri\} d^4x \nonumber\\
	\nonumber \\
	&=& \int \l\{-R n_0\l[mc^2 + {\mathcal E}\l(\f{R n_0}{J}\ri)\ri] + \ri.\nonumber\\
	& & \hspace{4em}\l.\f{R^2}{2J} b_0^i b_0^j\l( \pp{q^\lambda}{a^i} - \dot{q}^\lambda\dot{q}_\mu\pp{q^\mu}{a^i} \ri)\l( \pp{q_\lambda}{a^j} - \dot{q}_\lambda\dot{q}^\zeta\pp{q_\zeta}{a^j} \ri)\ri\} d^4a \nonumber\\
	\nonumber \\
	&=:& \int {\mathcal L}_\mathrm{MHD} d^4a.
\label{e:MHD Lagrangian}
\end{eqnarray}
Here, $\dot{q}^\lambda\dot{q}_\lambda$ in the third term will be evaluated as 1 after taking a variation.
The variation of ${\mathcal L}_\mathrm{MHD}$ with respect to $q^\nu$ gives the equation of the motion as
\begin{eqnarray*}
  & & \pp{}{s}(h\dot{q}_\nu) - \f{1}{ n_0}\pp{}{s}\l\{ \l[\f{1}{J} b_0^i b_0^j\l( \pp{q^\lambda}{a^i}\pp{q_\lambda}{a^j} - \dot{q}^\lambda\dot{q}_\mu\pp{q^\mu}{a^i}\pp{q_\lambda}{a^j} \ri) \ri]\dot{q}_\nu \ri\} \nonumber\\
	\nonumber\\
	&-& \f{1}{ n_0}C_\nu^{\;\;\mu}\pp{p}{a^\mu} + \f{1}{ n_0}C_\nu^{\;\;\mu}\pp{}{a^\mu}\l[\f{1}{2J^2} b_0^i b_0^j\l( \pp{q^\lambda}{a^i}\pp{q_\lambda}{a^j} - \dot{q}^\lambda\dot{q}_\mu\pp{q^\mu}{a^i}\pp{q_\lambda}{a^j} \ri) \ri] \nonumber\\
	\nonumber\\
	&-& \f{1}{ n_0}\pp{}{a^i}\l[ \f{1}{J} b_0^i b_0^j\l( \pp{q_\nu}{a^j} - \dot{q}^\mu\dot{q}_\nu\pp{q_\mu}{a^j} \ri) \ri] \nonumber\\
	\nonumber\\
	&+& \f{1}{ n_0}\pp{}{s}\l[ \f{1}{J} b_0^i b_0^j\l( \dot{q}_\mu\pp{q^\mu}{a^i}\pp{q_\nu}{a^j} - \dot{q}_\nu\dot{q}_\mu\dot{q}^\lambda\pp{q^\mu}{a^i}\pp{q_\lambda}{a^j} \ri) \ri] = 0 . \nonumber\\
\end{eqnarray*}
Using the identity (\ref{e:d/da = J d/dx(J dq/da)}), the second term on the left hand side is manipulated as,
\begin{eqnarray*}
	 - \f{J}{n_0}\pp{}{x^\zeta}\l\{ \l[\f{1}{J^2} b_0^i b_0^j\l( \pp{q^\lambda}{a^i}\pp{q_\lambda}{a^j} - \dot{q}^\lambda\dot{q}_\mu\pp{q^\mu}{a^i}\pp{q_\lambda}{a^j} \ri) \ri]\dot{q}_\nu\dot{q}^\zeta \ri\}\\
	 \nonumber \\
	 = - \f{1}{ n}\pp{}{x^\zeta}(b^2U^\zeta U_\nu),
\end{eqnarray*}
and the fifth and sixth terms are manipulated as,
\begin{eqnarray*}
	-\f{J}{ n_0}\pp{}{x^\zeta}\l\{ 
		  \f{1}{J^2} b_0^i b_0^j\l( \pp{q_\nu}{a^j} - \dot{q}^\mu\dot{q}_\nu\pp{q_\mu}{a^j} \ri)\pp{q_\zeta}{a^i} \ri. \\
			\hspace{7em}\l.- \f{1}{J^2} b_0^i b_0^j\l( \dot{q}_\mu \pp{q^\mu}{a^i} \pp{q_\nu}{a^j} - \dot{q}_\nu\dot{q}_\mu\dot{q}^\lambda\pp{q^\mu}{a^i}\pp{q_\lambda}{a^j} \ri)\dot{q}_\zeta
	\ri\} \\
	\\
	= -\f{1}{ n}\p_\lambda(b^\lambda b_\nu).
\end{eqnarray*}
Using (\ref{e:A d/da = J d/dx}), the third and fourth terms become,
\begin{eqnarray*}
	- \f{1}{n}\pp{p}{x^\nu} + \f{1}{n}\pp{}{x^\nu}\l(\f{b^2}{2}\ri).
\end{eqnarray*}
Then we obtain the Eulerianized equation (\ref{e:Eulerian MHD e.o.m.}).

%
%
\section{Relabeling symmetry and the relativistic helicity}\label{s:R-Helicity conservation via Noether}

\subsection{Relativistic helicity in plasma}\label{ss:Noether PL}
In this section we derive the conservation of the relativistic canonical helicity in plasma from Noether's theorem.
Let us suppose the label variation while the field variables are kept constant,
\begin{eqnarray*}
	\Delta q^\nu = 0 \;\;\;\mr{and}\;\;\; \delta a^\nu \ne 0,
\end{eqnarray*}
resulting
\begin{eqnarray*}
	\Delta \l( \pp{q^\nu}{a^\mu} \ri) &=& -\pp{(\delta a^\lambda)}{a^\mu}\pp{q^\nu}{a^\lambda}, \\
	\delta q^\nu        &=& -\delta a^\lambda\pp{q^\nu}{a^\lambda}.
\end{eqnarray*}
These are called as the relabeling transformation.
We investigate $\delta a^\nu$ which does not alter the plasma action (\ref{e:PL Lagrangian}).
From (\ref{e:variation of PL action}), we find
\begin{eqnarray*}
	\pp{{\mathcal L}_\mathrm{PL}}{\dot{q}^\nu} = - n_0 (h\dot{q}_\nu + eA_\nu) + pC_\nu^{\;\;0} \;\;\;\mr{and}\;\;\;
	\pp{{\mathcal L}_\mathrm{PL}}{(\tilde{\p}_i q^\nu)} = pC_\nu^{\;\;i},
\end{eqnarray*}
where $\tilde{\p}_\mu = \p/\p a^\mu$.
Let us define $\omega^i :=  n_0 \delta a^i$.
We obtain
\begin{equation*}
	\mathcal{L}_\mathrm{PL}\delta a^\mu + \delta q^\nu\pp{\mathcal{L}_\mathrm{PL}}{(\tilde{\p}_\mu q^\nu)} = \l( {\mathcal P}_\nu \omega^i\pp{q^\nu}{a^i},\; -(h + e\varrho)\bm{\omega} \ri).
\end{equation*}
Thereby we manipulate the integrand in the right-hand side of (\ref{e:delta S}) as
\begin{eqnarray}
	& & \pp{}{a^\mu}\l( \mathcal{L}_\mathrm{PL}\delta a^\mu + \delta q^\nu\pp{\mathcal{L}_\mathrm{PL}}{(\tilde{\p}_\mu q^\nu)} \ri) \nonumber\\
	\nonumber\\
	&=& \dot{{\mathcal P}}_\nu \omega^i \pp{q^\nu}{a^i} + {\mathcal P}_\nu \dot{\omega}^i \pp{q^\nu}{a^i} + \pp{}{a^i}\l[ - (h + e\varrho)\omega^i \ri] \nonumber\\
	\nonumber\\
	&=& \pp{(h + e\varrho - \theta)}{a^i}\omega^i + {\mathcal P}_\nu \dot{\omega}^i \pp{q^\nu}{a^i} + \pp{}{a^i}\l[ - (h + e\varrho)\omega^i \ri],\nonumber\\
	\label{e:div(I - delta lambda)}
\end{eqnarray}
where we used (\ref{e:Lagrangian e.o.m.}) and (\ref{e:thermodynamic relation}).
If the label transformation satisfies
\begin{equation}
	\pp{\omega^i}{a^i} = 0, 
\label{e:PL Noether condition1}
\end{equation}
\begin{equation}
	\dot{\omega}^i = 0, 
\label{e:PL Noether condition2}
\end{equation}
the right-hand side of (\ref{e:div(I - delta lambda)}) becomes $-\p(\theta\omega^i)/\p a^i$.
Therefore $\delta \Lambda^\mu = (0,\, -\theta\omega^i)$, and a Noether current is given as
\begin{equation}
	I^\mu = \l( {\mathcal P}_\nu \omega^i\pp{q^\nu}{a^i},\; -(h + q\varrho - \theta)\bm{\omega} \ri).
\label{e:PL I}
\end{equation}
The conservation of the Noether current ($\p I^\mu/\p a^\mu = 0$) is rewritten as
\begin{equation}
	\pp{}{s}(\bm{{\mathscr P}} \cdot \bm{\omega}) + \nbl_a\cdot\l[ -(h + q\varrho - \theta)\bm{\omega} \ri] = 0,
	\label{e:general Lagrangian conservation}
\end{equation}
where ${\mathscr P}_i := {\mathcal P}_\mu\p q^\mu/\p a^i$.

We define antisymmetric field tensor
\begin{eqnarray}
	{\mathcal W}^{*\mu\nu} := \f{\omega^i}{J}\l( \dot{q}^\mu\pp{q^\nu}{a^i} - \dot{q}^\nu\pp{q^\mu}{a^i} \ri).
\label{e:def W*}
\end{eqnarray}
By the use of (\ref{e:PL Noether condition1}), we can calculate
\begin{eqnarray}
	\p_\mu {\mathcal W}^{*\mu\nu} = 0,
\label{e:div W}
\end{eqnarray}
and (\ref{e:PL Noether condition2}) leads
\begin{equation}
	\p_\lambda(U^\lambda {\mathcal W}^{*\mu\nu}) = {\mathcal W}^{*\lambda\nu}\p_\lambda U^\mu + {\mathcal W}^{*\mu\lambda}\p_\lambda U^\nu.
\label{e:dot W}
\end{equation}
In terms of differential forms, if we consider ${\mathcal W}^{*\mu\nu}$ as the dual of 2-form ${\mathcal W}$, (\ref{e:div W}) corresponds to
\begin{equation}
	d{\mathcal W} = 0,
\label{e:dW}
\end{equation}
and (\ref{e:dot W}) corresponds to
\begin{equation}
	d i_U {\mathcal W} = 0.
\label{e:di_UW}
\end{equation}
The divergence-freeness of $\omega$ leads exactness of ${\mathcal W}$, and
the constancy of $\omega$ along the streamline leads transport of ${\mathcal W}$ by $U$, respectively.
Therefore the antisymmetric tensor ${\mathcal W}$ is ``vorticity like'', and we find (\ref{e:def W*}) is the Eulerianization to 2-form field.

\begin{figure}[htpb]
	\begin{center}
		\includegraphics*[width=0.8\textwidth]{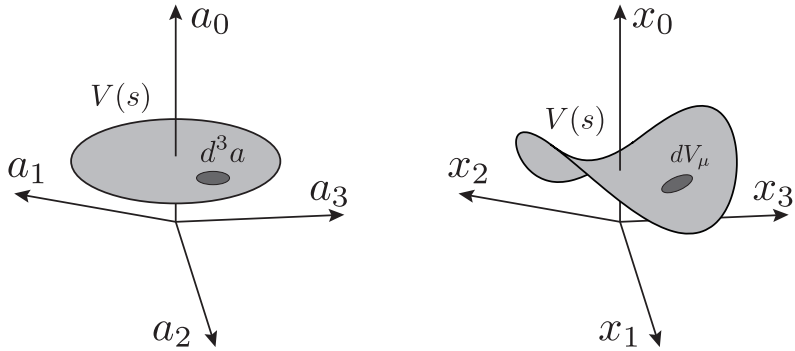}
	\end{center}
	\caption{Volume element transformation between the Lagrangian coordinates and the Eulerian coordinates}
	\label{f:volumes}
\end{figure}

Next we show the conservation of generalized helicity (\ref{e:canonical helicity conservation}) from the conservation of the Noether current $\p I^\mu/\p a^\mu = 0$.
Substituting (\ref{e:PL I}) and multiplying the volume element $da^1 \wedge	da^2 \wedge	da^3$, we calculate 
\begin{eqnarray*}
	\pp{I^\mu}{a^\mu} da^1 \wedge	da^2 \wedge	da^3 \nonumber\\
	\nonumber\\
	= \l\{ \pp{}{a^0}\l( {\mathcal P}_\nu\omega^i\pp{q^\nu}{a^i} \ri) + \pp{}{a^i}\l[ -(h + e\varrho - \theta)\omega^i \ri] \ri\}da^1 \wedge	da^2 \wedge	da^3 \nonumber\\
	\nonumber\\
	= C^{\;\;0}_\lambda\f{\dot{q}^\lambda}{J}\l\{ \pp{}{a^0}\l( {\mathcal P}_\nu\omega^i\pp{q^\nu}{a^i} \ri) \ri. \nonumber\\
	\hspace{9em}\l.+ \pp{}{a^i}\l[ -(h + e\varrho - \theta)\omega^i \ri] \ri\}da^1 \wedge	da^2 \wedge	da^3. \nonumber\\
\end{eqnarray*}
The purely spatial volume element $da^1 \wedge	da^2 \wedge	da^3$ in Lagrangian coordinates is mapped to a space-time mixed three-dimensional volume element $dV_\lambda$ embedded in the four-dimensional Eulerian coordinates (Fig.~\ref{f:volumes}): 
\begin{eqnarray}
	& & da^1 \wedge	da^2 \wedge	da^3 \nonumber\\
	\nonumber\\
	&=&\hspace{1em} \pp{(a^1,a^2,a^3)}{(x^1,x^2,x^3)}dx^1 \wedge dx^2 \wedge dx^3 + \pp{(a^1,a^2,a^3)}{(x^0,x^3,x^2)}dx^0 \wedge dx^3 \wedge dx^2 \nonumber\\
	\nonumber\\
	& &+ \pp{(a^1,a^2,a^3)}{(x^0,x^1,x^3)}dx^0 \wedge dx^1 \wedge dx^3 + \pp{(a^1,a^2,a^3)}{(x^0,x^2,x^1)}dx^0 \wedge dx^2 \wedge dx^1 \nonumber\\
	\nonumber\\
	&=& [C^{-1}]^{\;\;\lambda}_0\;dV_\lambda.
\label{e:d^3a to dV_mu}
\end{eqnarray}
Then we obtain
\begin{equation}
	\f{\dot{q}^\lambda}{J}\l\{ \pp{}{a^0}\l( {\mathcal P}_\nu\omega^i\pp{q^\nu}{a^i} \ri) + \pp{}{a^i}\l[ -(h + e\varrho - \theta)\omega^i \ri] \ri\}dV_\lambda = 0.
\label{e:L_U P^W4}
\end{equation}
Further manipulation leads
\begin{eqnarray}
	\l[ \p_\nu(U^\mu {\mathcal P}_\mu){\mathcal W}^{*\lambda\nu} + U^\mu(\p_\mu {\mathcal P}_\nu - \p_\nu {\mathcal P}_\mu){\mathcal W}^{*\lambda\nu} \ri.\nonumber\\
	\hspace{12em}\l. - \p_\nu(h + e\varrho - \theta){\mathcal W}^{*\lambda\nu} \ri]dV_\lambda = 0, 
\label{e:L_U P^W2}
\end{eqnarray}
which is the component representation of
\begin{equation}
	L_U({\mathcal P} \wedge {\mathcal W}) - d\l[ (h + e\varrho - \theta){\mathcal W} \ri] = 0.
\label{e:L_U P^W}
\end{equation}
By integrating we get
\begin{equation}
	\dd{}{s}\int_{V(s)}{\mathcal P} \wedge {\mathcal W} = 0.
  \label{e:generalized helicity conservation}
\end{equation}
Here ${\mathcal W}$ is any 2-form field which satisfies (\ref{e:dW}) and (\ref{e:di_UW}).
The arbitrariness of ${\mathcal W}$ is originated from that any label transformation satisfying (\ref{e:PL Noether condition1}) and (\ref{e:PL Noether condition2}) makes the action invariant.
In non-relativistic case, the arbitrariness in the helicity is already known~\cite{Fukumoto}.
Specifying as 
\begin{equation*}
	\omega^i = \epsilon^{0ijk}\pp{{\mathcal P}_\zeta}{a^j}\pp{q^\zeta}{a^k},
\end{equation*}
(\ref{e:generalized helicity conservation}) becomes (\ref{e:canonical helicity conservation}).

\subsection{Relativistic cross helicity in MHD}\label{ss:Noether MHD}
We apply the relabeling transformation to relativistic MHD Lagrangian (\ref{e:MHD Lagrangian}).
We can calculate
\begin{eqnarray}
	& & \mathcal{L}_\mathrm{MHD}\delta a^0 + \delta q^\nu\pp{\mathcal{L}_\mathrm{MHD}}{\dot{q}^\nu} \nonumber\\
	&=& \l[  n_0 P_\nu - \f{1}{J} b_0^i b_0^j\l( \pp{q^\lambda}{a^i}\pp{q_\lambda}{a^j} - \dot{q}^\lambda\dot{q}_\mu\pp{q^\mu}{a^i}\pp{q_\lambda}{a^j} \ri)\dot{q}_\nu \ri] \delta a^k \pp{q^\nu}{a^k} \nonumber\\
	& &\hspace{6em}+ \f{1}{J} b_0^i b_0^j\l( \dot{q}_\mu\pp{q^\mu}{a^i}\pp{q_\nu}{a^j} - \dot{q}_\nu\dot{q}_\mu\dot{q}^\lambda\pp{q^\mu}{a^i}\pp{q_\lambda}{a^j} \ri)\delta a^k \pp{q^\nu}{a^k} \nonumber\\
	\nonumber\\
	& & \mathcal{L}_\mathrm{MHD}\delta a^k + \delta q^\nu\pp{\mathcal{L}_\mathrm{MHD}}{(\tilde{\p}_k q^\nu)} \nonumber\\
	&=& -\l[  n_0 h - \f{1}{J} b_0^i b_0^j\l( \pp{q^\lambda}{a^i}\pp{q_\lambda}{a^j} - \dot{q}^\lambda\dot{q}_\mu\pp{q^\mu}{a^i}\pp{q_\lambda}{a^j} \ri) \ri] \delta a^k \nonumber\\
	& & \hspace{10em}- \f{1}{J} b_0^j b_0^k\l( \pp{q_\nu}{a^j} - \dot{q}^\mu\dot{q}_\nu\pp{q_\mu}{a^j} \ri)\delta a^i \pp{q^\nu}{a^i}. \nonumber\\
	\label{e:MHD I - delta lambda}
\end{eqnarray}
Thereby we manipulate the integrand in the right-hand side of (\ref{e:delta S}) as
\begin{eqnarray}
	\pp{}{a^\mu}\l( \mathcal{L}_\mathrm{MHD}\delta a^\mu + \delta q^\nu\pp{\mathcal{L}_\mathrm{MHD}}{(\tilde{\p}_\mu q^\nu)} \ri) \nonumber\\
	\nonumber\\
	= -\pp{}{a^i}( \theta \omega^i) + \f{1}{J}\l[ \pp{ b_0^i}{a^k} b_0^j\f{\omega^k}{n_0} + b_0^i b_0^j \pp{}{a^k}\l( \f{\omega^k}{n_0} \ri) \ri.\nonumber\\
	\hspace{7em}\l.- b_0^k b_0^j \pp{}{a^k}\l( \f{\omega^i}{n_0} \ri) \ri]\l( \pp{q^\lambda}{a^i}\pp{q_\lambda}{a^j} - \dot{q}^\lambda\dot{q}_\mu\pp{q^\mu}{a^i}\pp{q_\lambda}{a^j} \ri). \nonumber\\
	\label{e:MHD div(I - delta lambda)2}
\end{eqnarray}
For this to be exact differential, the condition for the label transformation (\ref{e:PL Noether condition1}) and (\ref{e:PL Noether condition2}) are not sufficient; additionally we require
\begin{equation}
	 \omega^i =  b_0^i.
\label{e:MHD-Noether condition3}
\end{equation}
This means that the degree of freedom in the relabeling symmetry is reduced;
Plugging (\ref{e:MHD-Noether condition3}), the right-hand side of (\ref{e:MHD div(I - delta lambda)2}) becomes $-\p(\theta b_0^i)/\p a^i$.
Therefore $\delta \Lambda^\mu = (0,\, -\theta b_0^i)$, and a Noether current is given as
\begin{equation}
	I^\mu = \l( P_\mu  b_0^i \pp{q^\mu}{a^i},\, -(h - \theta)\bm{b}_0 \ri).
	\label{e:MHD I}
\end{equation}
Here the momentum is mechanical (not ${\mathcal P}$ but $P$). 
In the same way as the previous plasma case by replacing $\omega^i \to b_0^i$ and $e = 0$,
we can show the conservation of the Noether current $\p I^\mu/\p a^\mu = 0$ leads
\begin{equation*}
	\dd{}{s}\int_{V(s)}P \wedge {\mathcal F} = 0.
\end{equation*}
The generalized field ${\mathcal W}$ is specified as ${\mathcal F}$ in the definition of the MHD cross helicity because $\omega^i$ is specified as $ b_0^i$ in the Noether current (\ref{e:MHD I}) due to the additional term in the MHD Lagrangian (\ref{e:MHD Lagrangian}).

%
%
\section{Summary}
The conservation of the relativistic helicity is derived from Noether's theorem pertaining to the fluid elements' relabeling symmetry. 
In plasma, labeling transformation has freedom to some extent; any transformation satisfying (\ref{e:PL Noether condition1}) and (\ref{e:PL Noether condition2}) preserve the plasma action. 
Therefore relativistic canonical helicity in plasma is generalized as the external product of the canonical momentum ${\mathcal P}$ and field tensor ${\mathcal W}$ which satisfies (\ref{e:dW}) and (\ref{e:di_UW}).
Then we discovered the Lagrangian description of the relativistic MHD.
The freedom of labeling transformation is reduced in the MHD Lagrangian, then only transformation $b_0$ makes the action invariant.
Resulting helicity becomes the external product of the mechanical momentum $P$ and MHD tensor ${\mathcal F}$.

In non-relativistic dynamics, the time in Lagrangian and Eulerian coordinates coincides.
Therefore the divergence free current $(\p I^\mu/\p a^\mu)da^1 \wedge da^2 \wedge da^3 = 0$ in the Lagrangian coordinates can be transformed into some divergence free current $(\p K^\mu/\p x^\mu)dx^1 \wedge dx^2 \wedge dx^3 = 0$ in the Eulerian coordinate~\cite{Padhye-Morrison1,Padhye-Morrison2}.
On the other hand, in relativistic dynamics, the time on the Lagrangian coordinates becomes proper time $s$ because the time is measured on the co-moving fluid element
while the time on the Eulerian coordinates is reference time $t$.
As we discovered in the present work, due to the difference of $s$ and $t$, Eulerianized conservation law (\ref{e:L_U P^W}) is no longer a divergence of some current;   
Technically we require (\ref{e:d^3a to dV_mu}) for transforming the conservation of the current into Eulerian coordinate because the Jacobian is defined as $d^4x = Jd^4a$ (while in non-relativistic case, $d^3x = Jd^3a$).
We must be careful when we use the term ``current'' in fluid/plasma theory. It should be used in Lagrangian coordinate. 
%

\ack
We appreciate Professor Philip J Morrison for helpful comments and discussions.  
The work of Y.K. was supported by Grant-in-Aid for JSPS Fellows 241010.

\appendix
%
%
\section{Identities of Jacobian and cofactor}\label{a:identities}
Here we show the four dimensional determinant identities, which is simple extension of the three dimensional case \cite{Padhye}.
We define $C_\mu^{\;\;\nu}$ as the cofactor of the matrix element $\p q^\mu/\p a^\nu$ which is expressed as
\begin{equation}
	C_\mu^{\;\;\nu} := \f{1}{6}\epsilon_{\mu\alpha\beta\gamma}\epsilon^{\nu\delta\sigma n}\pp{q^\alpha}{a^\delta}\pp{q^\beta}{a^\sigma}\pp{q^\gamma}{a^ n}.
\label{e:def A}
\end{equation}
The determinant $J$ is given as
\begin{equation}
	J\delta^\lambda_\mu = C_\mu^{\;\;\nu}\pp{q^\lambda}{a^\nu}.
\label{e:J and A1}
\end{equation}
Differentiating (\ref{e:J and A1}) by $\p q^\lambda/\p a^\nu$, 
\begin{equation*}
	C_\mu^{\;\;\nu} = \pp{J}{(\tilde{\p}_\mu q^\nu)}.
\end{equation*}
Next, differentiating (\ref{e:def A}) with respect to $a^\nu$, we obtain
\begin{equation}
	\pp{}{a^\nu}C_\mu^{\;\;\nu} = 0.
\label{e:dA/da}
\end{equation}
This leads
\begin{equation}
	C_\mu^{\;\;\nu}\pp{f}{a^\nu} = C_\mu^{\;\;\nu}\pp{q^\lambda}{a^\nu}\pp{f}{x^\lambda} = J\pp{f}{x^\mu},
\label{e:A d/da = J d/dx}
\end{equation}
and
\begin{equation}
	\pp{f}{a^\nu} = J\pp{}{x^\mu}\l( \f{1}{J}\pp{q^\mu}{a^\nu}f \ri),
\label{e:d/da = J d/dx(J dq/da)}
\end{equation}
where $f$ is some function.
We contract (\ref{e:J and A1}) with $\p[q^{-1}]^\sigma/\p x^\lambda$ then obtain
\begin{equation}
	J \pp{[q^{-1}]^\sigma}{x^\mu} = C_\mu^{\;\;\sigma}.
\label{e:J and A2}
\end{equation}
Contracting (\ref{e:J and A2}) with $\p q^\mu/\p a^\nu$ gives
\begin{equation}
	J \delta^\sigma_\nu = C_\mu^{\;\;\sigma}\pp{q^\mu}{a^\nu}.
\label{e:J and A3}
\end{equation}

Next contracting (\ref{e:J and A1}) with $\epsilon^{\mu\phi\psi\tau}$, we obtain
\begin{equation}
	J\epsilon^{\lambda\phi\psi\tau} = \epsilon^{\nu\delta\sigma n}\pp{q^\lambda}{a^\nu}\pp{q^\phi}{a^\delta}\pp{q^\psi}{a^\sigma}\pp{q^\tau}{a^ n}.
\label{e:J epsilon}
\end{equation}
Differentiating (\ref{e:J epsilon}) with respect to $s$ leads to
\begin{equation}
	\pp{J}{s} = C_\mu^{\;\;\nu}\pp{\dot{q}^\mu}{a^\nu}.
\label{e:dot J}
\end{equation}
This identity naturally induces the mass conservation law.
Multiplying $-J^{-2}$, 
\begin{equation}
	\pp{J^{-1}}{s} = -\f{1}{J}\f{C_\mu^{\;\;\nu}}{J}\pp{\dot{q}^\mu}{a^\nu} = -\f{1}{J}\pp{\dot{q}^\mu}{x^\mu}.
\label{e:continuity}
\end{equation}
Multiplying again $ n_0$ gives continuity equation $\p_\mu( n U^\mu) = 0$.
\\

%
%

\end{document}